\documentclass[sigconf]{acmart}

\usepackage{amsmath}
\usepackage{amsfonts}
\usepackage{tabularx}
\usepackage{algorithm,algpseudocode}
\usepackage{array,multirow}
\usepackage{xspace}
\usepackage{pifont}
\usepackage{adjustbox}
\usepackage{xcolor}
\usepackage{colortbl}
\usepackage{pbox}
\usepackage{graphicx}
\usepackage{subcaption}
\usepackage{tikz}

\newcommand{\ziniu}[1]{{{\color{red}(Ziniu: #1)}}}
\newcommand{\ours}{\textit{Stage}\xspace}
\long\def\comment#1{}

\AtBeginDocument{%
  }

\setcopyright{acmcopyright}
\copyrightyear{2024}
\acmYear{2024}
\acmDOI{XXXXXXX.XXXXXXX}

\acmConference[SIGMOD'24]{ACM SIGMOD/PODS International Conference on Management of Data}{June 09--15,
  2024}{Santiago, Chile}

\begin{document}

\title{Stage: Query Execution Time Prediction in Amazon Redshift}



\author{Ziniu Wu$^{\dagger}$}
\email{ziniuw@mit.edu}
\affiliation{
\institution{MIT CSAIL}
\country{USA}
}
\author{Ryan Marcus$^{\dagger}$}
\email{rcmarcus@seas.upenn.edu}
\affiliation{
\institution{University of Pennsylvania}
\country{USA}
}
\author{Zhengchun Liu}
\email{zcl@amazon.com}
\affiliation{
\institution{Amazon Web Services}
\country{USA}
}
\author{Parimarjan Negi$^{\dagger}$}
\email{pnegi@mit.edu}
\affiliation{
\institution{MIT CSAIL}
\country{USA}
}
\author{Vikram Nathan}
\email{vrnathan@amazon.com}
\affiliation{
\institution{Amazon Web Services}
\country{USA}
}
\author{Pascal Pfeil}
\email{pfeip@amazon.de}
\affiliation{
\institution{Amazon Web Services}
\country{Germany}
}
\author{Gaurav Saxena}
\email{gssaxena@amazon.com}
\affiliation{
\institution{Amazon Web Services}
\country{USA}
}
\author{Mohammad Rahman}
\email{rerahman@amazon.com}
\affiliation{
\institution{Amazon Web Services}
\country{USA}
}
\author{Balakrishnan (Murali) Narayanaswamy}
\email{muralibn@amazon.com}
\affiliation{
\institution{Amazon Web Services}
\country{USA}
}
\author{Tim Kraska$^{\dagger}$}
\email{kraska@mit.edu}
\affiliation{
\institution{Amazon Web Services, MIT CSAIL}
\country{USA}
}

\renewcommand{\shortauthors}{Wu et al.}


\begin{abstract}
Query performance (e.g., execution time) prediction is a critical component of modern DBMSes. As a pioneering cloud data warehouse, Amazon Redshift relies on an accurate execution time prediction for many downstream tasks, ranging from high-level optimizations, such as automatically creating materialized views, to low-level tasks on the critical path of query execution, such as admission, scheduling, and execution resource control. 
Unfortunately, many existing execution time prediction techniques, including those used in Redshift, suffer from cold start issues, inaccurate estimation, and are not robust against workload/data changes. 

In this paper, we propose a novel hierarchical execution time predictor: the \emph{Stage} predictor. The \emph{Stage} predictor is designed to leverage the unique characteristics and challenges faced by Redshift.
The Stage predictor consists of three model states: an execution time \emph{cache}, a lightweight \emph{local model} optimized for a specific DB instance with uncertainty measurement, and a complex \emph{global model} that is transferable across all instances in Redshift. 
We design a systematic approach to use these models that best leverages optimality (cache), instance-optimization (local model), and transferable knowledge about Redshift (global model).
Experimentally, we show that the Stage predictor makes more accurate and robust predictions while maintaining a practical inference latency and memory overhead.
Overall, the Stage predictor can improve
the average query execution latency by $20\%$ on these instances compared to the prior query performance predictor in Redshift.

\end{abstract}

\maketitle

\begingroup
\renewcommand\thefootnote{}\footnote{\noindent
	$\dagger$ Work conducted while affiliated with Amazon Web Services.
}
\addtocounter{footnote}{-1}
\endgroup

\vspace{-1em}
\section{Introduction}
\label{sec: intro}

Predicting the execution time (exec-time) of a query before actually executing the query is a crucial component for a number of tasks in intelligent cloud DBMSes, such as query optimization~\cite{SunL19, HilprechtB22}, workload scheduling~\cite{WagnerK021, PatelDZPZSMS18}, admission control~\cite{TozerBA10}, resource management~\cite{TaftLDESD16, LyuFSSDCMFLZZ22, ChenDX21}, and maintaining SLAs~\cite{ChiMHT11, MarcusP16}.

Amazon Redshift, a pioneering cloud data warehouse, relies on exec-time prediction for many downstream tasks, ranging from high-level optimizations (e.g., automatically creating materialized views~\cite{ArmenatzoglouBB22}) to low-level tasks on the critical path of query execution (e.g. admission, scheduling and execution resource control inside its workload manager~\cite{SaxenaRCLCCMKPN23}).
For example, the workload manager in Redshift separates queries into ``short-running'' and ``long-running'' queues based on estimated exec-time.  
The short-running query queue has its own dedicated resources and unique optimizations to meet users' expectations of fast execution. 
If a long-running query is erroneously placed into the short-running queue by the exec-time predictor, the long-running query can cause head-of-line blocking, significantly delaying the execution of short-running queries in the queue.
Conversely, if a shorting-running is wrongly placed into the long-running queue, the short query may queue up for minutes before execution. Both cases can severely affect the overall query performance on a cluster and the user experience of Redshift.

The existing exec-time predictor inside Amazon Redshift (AutoWLM predictor~\cite{SaxenaRCLCCMKPN23}) uses an instance-optimized XGBoost model~\cite{ChenG16} trained on each customer's database cluster, using each cluster's executed queries. This model is very lightweight to ensure negligible inference latency and memory overhead on the critical path of query execution. However, the AutoWLM predictor has the following downsides. 
First, due to its lightweight nature and simplified query featurization techniques, it can produce \emph{inaccurate estimations}. 
Second, whenever the customers' data or query workload changes, it can provide \emph{unreliable predictions} until the predictor's training set ``catches up'' with the change.
Third, the AutoWLM predictor requires a sufficient amount of executed queries as training examples, which may not be available for a new instance, and thus, it performs poorly in \emph{cold start} scenarios.

\begin{figure}
     \centering
     \begin{subfigure}{0.235\textwidth}
         \centering
         \includegraphics[width=\textwidth]{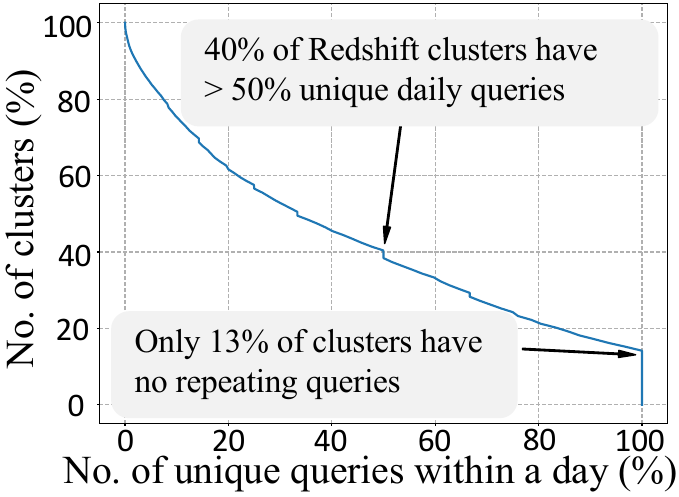}
         \caption{Distribution of clusters by the \% of queries that were unique within a day (not repeated).}
         \label{fig:repeat_dist}
     \end{subfigure}
     \hfill
     \begin{subfigure}{0.235\textwidth}
         \centering
         \includegraphics[width=\textwidth]{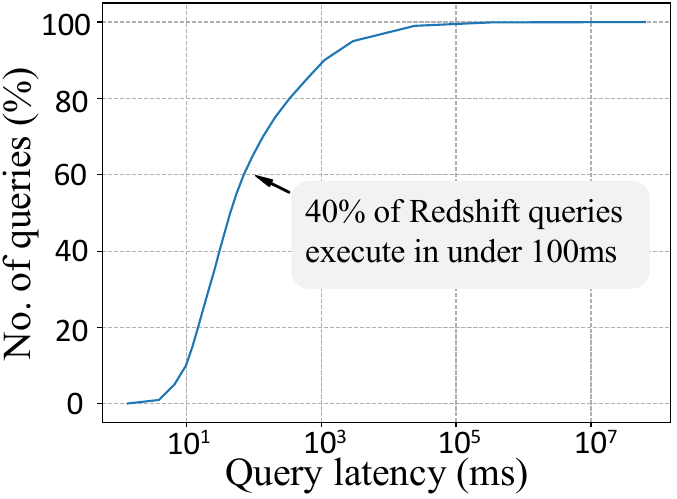}
         \caption{Distribution of query latency across the Redshift fleet (0.01\% to 99.99\% shown).}
         \label{fig:latency_dist}
     \end{subfigure}
     \vspace{-2em}
     \caption{Distribution statistics from the Redshift fleet}
     \vspace{-0em}
     \label{fig:three graphs}
\end{figure}

We make two key observations about the Amazon Redshift fleet that motivate the design of our new exec-time predictor. First, most queries executed on Amazon Redshift are \emph{low latency queries}. 
Many queries execute in just a few milliseconds, so naively applying the advanced exec-time predictors in recent literature~\cite{SunL19, MarcusP19, HilprechtB22}, which have inference time on the order of 50ms to 500ms, on the critical path will result in more time being spent on prediction than on actual query execution.
Thus, despite the superior estimation accuracy of these modern techniques, their inference latency overhead is not affordable for a lot of Redshift queries.
For example, the predictor proposed in~\cite{MarcusP19} has a prediction latency of 100ms, which is longer than $40\%$ of queries running in Redshift (as shown in Figure~\ref{fig:latency_dist})!
Second, Amazon Redshift customers tend to issue \emph{repeating queries}. On average, more than $60\%$ of the queries executed within Amazon Redshift have been executed within 24 hours of the execution of an identical query (as shown in Figure~\ref{fig:repeat_dist}).\footnote{These queries are \emph{exactly} repeated, both in terms of SQL and parameter values, but the database may have changed in the meantime. Note queries served by the Amazon Redshift result cache, which caches the results of repetitive queries when the underlying database has not changed, are not included in Figure~\ref{fig:repeat_dist}.}

\begin{figure}[t]
    \centering
    \includegraphics[width=8cm]{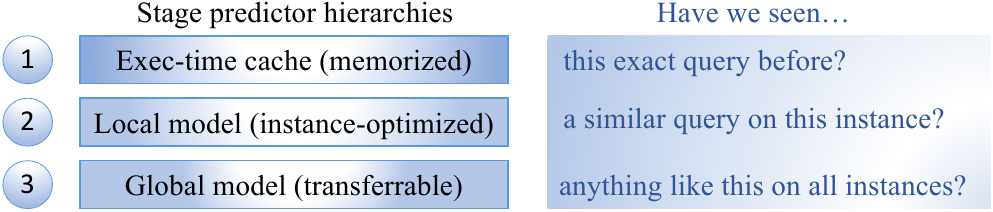}
   	\vspace{-1em}
    \caption{The key components and ideas of \ours predictor.}
    \vspace{-1em}
    \label{fig:motivation}
\end{figure}

To address the challenges of cold-start prediction, inference time, and reliable estimation, we implemented a novel hierarchical exec-time predictor (\ours) with three stages of models illustrated in Figure~\ref{fig:motivation}: (1) a local exec-time cache, which simply memorizes the latency of recently executed queries and predicts that latency when the exact query is submitted again, (2) a local lightweight exec-time predictor with uncertainty measurement that is instance-optimized to each Redshift customer, and (3) a complex global predictor that is transferable across all instances in Redshift.
When a new query $Q$ arrives, \ours predictor will first look up $Q$ in \textit{exec-time cache} and directly return its prediction based on previously observed exec-time if $Q$ is present in the cache.
If a query $Q$ misses the cache, \ours predictor will use a lightweight local exec-time predictor (\textit{local model}) to predict its exec-time and an uncertainty measurement of the prediction. The \textit{local model} utilizes a Bayesian ensemble of lightweight XGBoost models~\cite{MalininPU21} that can 
provide a query exec-time prediction and a reliable uncertainty measure associated with the prediction with a very low inference latency. 
While the \textit{local model} never learns a fully generalizable model of query performance, it can accurately predict queries similar to the past-seen queries. Thus, it can be thought of as a ``fuzzy cache''.
The prediction uncertainty can be high whenever the \textit{local model}  does not have enough training examples, or the input query is very different from previously seen queries.
In this situation, \ours predictor will use the \textit{global model}. Inspired by the recent advance in zero-shot cost model~\cite{HilprechtB22}, we design our \textit{global model} as a graph neural network~\cite{0004JWLJJH22, scarselli2008graph} that takes a physical execution plan of a query as input to predict its exec-time. 
There exist tens of thousands of instances in Redshift with diversified workloads. 
Thus, we train a single \textit{global model} on a diverse set of instances to distill the transferable knowledge of exec-time prediction across various instances. As a result, it is capable of accurately and robustly predicting the exec-time of queries on unseen clusters. 
The \textit{global model} will have a non-trivial inference latency (up to 100ms). Therefore, when used on a critical path of query execution, it will only be used when the \textit{local model} is uncertain about its prediction and believes the query's exec-time to be longer than a couple of seconds. Because the \textit{global model} is rarely used, the additional inference overhead is amortized out.

We simulate \ours predictor inside the workload manager~\cite{SaxenaRCLCCMKPN23} of Amazon Redshift in an actual production environment.
We conduct end-to-end evaluations on the 100 most billed instances in the month of July 2023 for each of three regions: `us-east-1', `us-west-2', and `eu-west-1'.
The results show that \ours predictor can improve the average query execution latency by $20\%$ on these instances compared to the prior exec-time predictor in Redshift.
In addition, we conduct thorough ablation studies to demonstrate the performance and reliability of each component of \ours predictor. 

To the best of our knowledge, \ours predictor is the first to apply the idea of a hierarchy of models in exec-time prediction or similar tasks in DBMS. 
We believe that \ours predictor points out a way to practically integrate expensive machine learning models on the critical path of customer-facing production systems. We list the main contributions of this paper as follows:

\begin{itemize}
    \item We describe the use cases and unique challenges of the exec-time predictor in Redshift (Section~\ref{sec: background} and Section~\ref{sec:design_p}).
    \item We design a \ours predictor framework with hierarchical components for exec-time prediction (Section~\ref{sec: method}).
    \item We show a comprehensive evaluation of production data to showcase the advantages of \ours predictor (Section~\ref{sec: exp}).
    \item We summarize the lessons learned and point out important research questions (Section~\ref{sec: lessons}).
\end{itemize}


\section{Background}
In this section, we first give an overview of query processing in Amazon Redshift, and then we provide a brief survey of related works on query exec-time prediction.
\label{sec: background}

\subsection{Execution time predictor in Redshift}
\label{subsec: auto-wlm}

\begin{figure}[t]
    \centering
    \includegraphics[width=8cm]{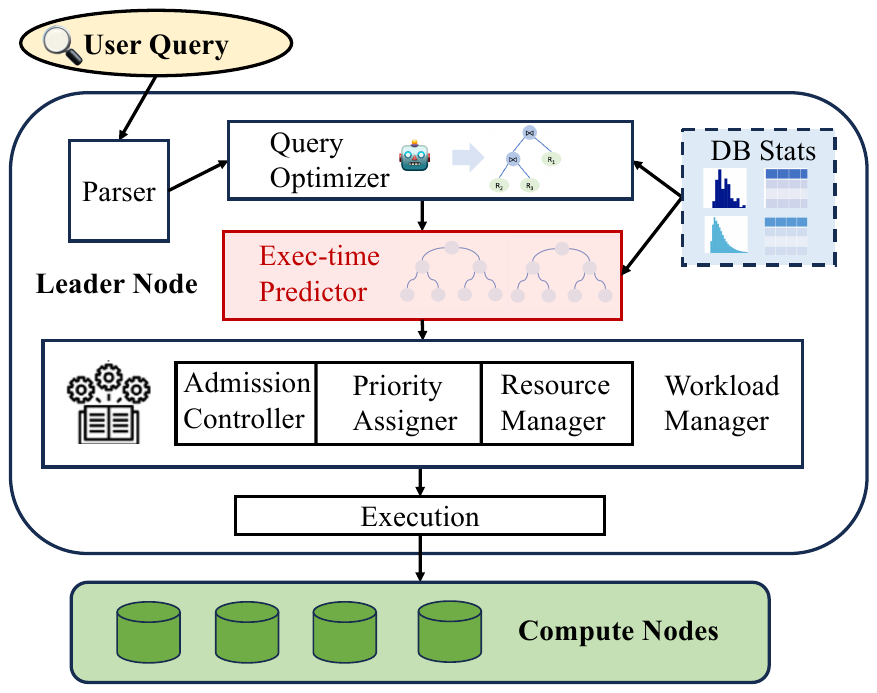}
    \vspace{-0.5em}
    \caption{Queries' lifetime inside Redshift.}
    \vspace{-1em}
    \label{fig: redshift-overview}
\end{figure}

Figure~\ref{fig: redshift-overview} illustrates the lifetime of user-issued queries inside Redshift. The queries will first go through a parser and query optimizer to derive their physical execution plans. Then, the exec-time predictor will take these plans as input and predict their exec-time. Based on their predictions, the workload manager will make a series of choices to determine their execution strategy and resource allocations (see~\cite{SaxenaRCLCCMKPN23} for an overview). Finally, the workload manager will send the queries out for actual execution.

\smallskip
\paragraph{Importance of exec-time predictor}
The predicted exec-time is a critical component of the workload manager's decision-making choices.
Based on the prediction, the admission controller in the workload manager will decide whether a query should wait in the queue, be pushed to a special short query queue, be executed on the user’s main cluster, or be sent to a concurrency scaling cluster.
For queries waiting in a queue, each query's priority is determined by the predicted exec-time (short queries execute first). 
If the workload manager decides to start up a new concurrency scaling cluster to process an incoming query, the optimal cluster size will be chosen based on the predicted exec-time on the candidate cluster sizes. 
Therefore, the accuracy of exec-time predictor directly affects query performance in Redshift. 
For example, if exec-time predictor is inaccurate, it can mistake a long-running query as short-running. This long-running query can block the execution of other short-running queries in the queue, thus severely degrading the overall query latency. Conversely, if a shorting-running is mistaken as long-running, it may queue up for minutes before execution.

In addition to the usage of exec-time predictor on the critical path of query execution, it is also used in several other high-level optimization tasks. 
For example, automatic materialized view creation in Redshift~\cite{ArmenatzoglouBB22} uses the query optimizer to regenerate queries' execution plans as if certain materialized view exists and then uses the exec-time predictor to estimate the performance of these plans to determine the benefits of building such materialized view. 

\paragraph{The prior AutoWLM exec-time predictor in Redshift}
Here, we summarize the current status of the exec-time predictor in Redshift as described in prior work~\cite{SaxenaRCLCCMKPN23}, the AutoWLM predictor. 
First, the AutoWLM predictor takes a physical execution plan of a query as input and flattens it into a vector. 
Then, a lightweight XGBoost model~\cite{ChenG16} is used to predict the query's exec-time.
As queries are executed in each instance, their feature vector and observed exec-time are added to the XGBoost model's training set.

The AutoWLM predictor is lightweight to ensure negligible inference latency and memory overhead on the critical path of query execution. 
However, due to its lightweight nature and simplified query functionalization techniques, it can sometimes produce inaccurate estimations. 
Worse yet, some customers’ data and query workload change quickly, thus making the predictions unreliable. 
In addition, the AutoWLM predictor requires sufficient executed
queries as training examples, which may not be available for a new instance (and hence a cold-start problem).
Moreover, many downstream tasks require not just an estimate of exec-time but also a confidence interval around that estimate for robust optimizations. For example, the automatic materialized view creation and cluster scaling model in Redshift need a confidence interval to ensure good worst-case behavior of the changes in the cluster. The AutoWLM predictor provides these confidence intervals using simple global statistics, which leaves room for improvement.

\subsection{Related Works}
\label{subsec: related}

\comment{
\begin{itemize}
    \item Cite a bunch of paper on XGBoost
    \item Explain the XGBoost with Bayesian ensemble paper
\end{itemize}

\textbf{Works on instance-optimized exec-time predictor (local model)} 

Briefly explain and say why it is impractical to use in production (easy to address: inference time too high)

\textbf{Carsten's work (zero-shot model)}

\begin{itemize}
    \item Explain his work 
    \item Discuss the benefits of such global model inside Redshift
    \item Discuss its impracticality and the challenges of implementing that in Redshift
\end{itemize}
}

Here, we give a brief overview of previous results in the areas of uncertainty quantification and query performance prediction.

\paragraph{Uncertainty of XGBoost models}
XGBoost is a scalable approach for building gradient boosting tree models, which achieved state-of-art performance in a wide range of tasks~\cite{ChenG16, OgunleyeW20, ZhangQMHHS18, TorlayPTB17}.
Recent work proposed a Bayesian ensemble of gradient boosting tree models to estimate the uncertainty of model prediction~\cite{MalininPU21}. 
We abuse the term XGBoost models to refer to gradient-boosting tree models for easier understanding.
In a nutshell, this approach separates the uncertainty into model and data uncertainty. 
It trains the XGBoost models with a probabilistic likelihood loss function~\cite{ProkhorenkovaGV18}. 
Thereafter, instead of predicting a single name, the XGBoost models will output a mean $\mu$ and variance $\sigma$ for its prediction, where $\mu$ is the model prediction and $\sigma$ captures the data uncertainty.
The Bayesian ensemble of XGBoost models independently trains several XGBoost models, denoted as $M_1, \ldots, M_k$, each of which will produce a $\mu_i$ and $\sigma_i$. The variance of all output means $\mu_1, \ldots, \mu_k$ captures the model uncertainty.
Finally, the total uncertainty of prediction is the sum of model uncertainty and data uncertainty.
\ours predictor adapts this approach to build an instance-optimized local predictor. The details will be discussed in Section~\ref{subsec: local}.

\paragraph{Instance-optimized exec-time predictor}
Traditional exec-time predictors generally use hand-derived heuristics and statistical models to understand the relational operators~\cite{DugganPCU14, WuCZTHN13, AkdereCRUZ12, LiKNC12}.
There has been a line of work using machine learning models to predict the exec-time of a query with superior accuracy over their traditional counterparts. 
In general, they featurize the logical or physical query plan as a graph and train graph neural networks to map the query plan to its exec-time~\cite{MarcusP19, SunL19, neo, bao, balsa, zhou2020query}.
One common drawback of these approaches is their high inference latency, preventing Redshift from integrating them on the critical path of query execution. Many Redshift queries execute in just a few milliseconds, and the inference latency of these methods surpasses a large proportion of queries’ actual execution time.

\paragraph{Zero-shot exec-time predictor}
In contrast to instance-optimization, zero-shot exec-time predictor~\cite{HilprechtB22} proposed to train one model over a diverse set of DB instances, and it can be directly used to predict the query exec-time on arbitrary unseen DB instances. 
Specifically, the zero-shot model gathers data-specific statistics from each DB instance, such as the number of tuples/columns/pages of each table.
Then, it embeds these statistics into a physical query execution plan to predict its exec-time. 
After a heavy offline training process, the zero-shot model understands the data-independent knowledge about the system and is transferrable to unseen DB instances. 
\ours predictor adapts this approach to build an instance-optimized global predictor inside Redshift. The details will be discussed in Section~\ref{subsec: global}.

\paragraph{Machine Learning for Databases}
In addition to exec-time prediction, machine learning has a large impact on optimizing and managing modern database systems. Machine learning systems help enable more effective and automated workload management~\cite{MarcusP16, SaxenaRCLCCMKPN23, buffer_sched}, index recommendation~\cite{DingDM0CN19}, and configuration tuning~\cite{AkenPGZ17}.
Machine learning algorithms help build more fine-grained and instance-optimized sub-components embedded in existing DBMSes, such cardinality estimation~\cite{HilprechtSKMKB20, YangKLLDCS20, ZhuWHZPQZC21, WuNAKM23, NegiWKTMMKA23}, learned query optimization~\cite{neo, bao, balsa, fastgres, skinnerdb}, learned indexes~\cite{KraskaBCDP18, KipfMRSKK020, NathanDAK20}, and learned storage layouts~\cite{DingNAK20, CurinoZJM10}.

\section{Design Principles}
\label{sec:design_p}

Redshift's query predictor has several design constraints, some likely to apply to other database engines, while others may be unique to Redshift. Here, we outline the three most important design principles behind our \ours exec-time predictor. 

First, like many OLAP databases, a large number of queries seen by Redshift are repeated queries (e.g., dashboard refreshes) -- taking advantage of this repetition is critical to correctly predicting latency for the majority of queries across the Redshift fleet. 
Second, standard point predictions (i.e., mean estimates) are insufficient for the predictor's downstream tasks; we require reasonable confidence bounds around each prediction to guarantee worst-case performance.
Third, the inference time of models on the critical path of query execution must be fast since a large portion of Redshift queries execute in only a few milliseconds. Thus, we set out to design a predictor with inference time on the order of microseconds.

\paragraph{Repeated queries} Many customers use Redshift for analytics tasks like dashboarding or report generation. As a result, identical queries are often repeatedly issued to Redshift. Figure~\ref{fig:repeat_dist} shows the distribution of the percentage of daily unique queries across the Redshift fleet: a ``daily unique'' query is a query sent to Redshift without an identical query being issued within the last 24 hours. Daily unique queries are thus a good \emph{lower bound} for the number of repeating queries that Redshift sees, as monthly or weekly reports will not appear as daily unique queries. We observe that past performance of a query is a strong predictor of the same query's performance later that day, since data distributions are \emph{normally} static day-by-day (distribution shifts do occur, but normally not within 24 hours). Therefore, we want to design our predictor to take advantage of these repeating queries. This motivates the first ``caching'' stage of our predictive model, discussed in Section~\ref{subsec: cache}.

\paragraph{High-confidence predictions} Many off-the-shelf machine learning models (e.g.,~\cite{scikit-learn}) give predictions as \emph{point estimates}, or approximations of the mean. However, Redshift uses predictions in a number of downstream tasks, including query scheduling and cluster sizing, so error bounds are essential for ensuring the entire system maintains good worst-case behavior. Error bounds are especially important for deciding when to dedicate more inference time to make a more accurate prediction, which we discuss in Section~\ref{subsec: local}.

\paragraph{Low inference latency} Since Redshift needs to estimate the exec-time for every issued query, it is important that the model's inference procedure is efficient. Figure~\ref{fig:latency_dist} shows the distribution of query latency across the Redshift fleet. Most Redshift queries execute in under $100ms$. 
This rules out exclusively using some modern advanced models, which could have inference times as high as $100ms$~\cite{MarcusP19, SunL19, HilprechtB22} (higher than the total query latency for 40\% of the queries!). For Redshift, our new \ours predictor only uses an expensive neural network model when we have high confidence that a query will be long (details in Section~\ref{subsec: global}). In this case, the additional inference time is a trivial portion of the overall exec-time.

\section{Stage Predictor}
\label{sec: method}

In order to meet the specific needs of Redshift, we design the \ours exec-time predictor to work in \emph{stages}. The first stage of the model, the \emph{exec-time cache} (Section~\ref{subsec: cache}), remembers the recently executed queries. When an incoming query matches a past query, the \emph{exec-time cache} makes a prediction. If a match is not found, the query proceeds to the next stage, the \emph{local model} (Section~\ref{subsec: local}). The \emph{local model} is instance-optimized to each user's clusters (i.e., trained per cluster). While the \emph{local model} never learns a fully generalizable model of query performance, it can accurately predict queries that are slight modifications of past-seen queries. Thus, it can be thought of as a ``fuzzy cache''.
When it cannot make a confident prediction, the query proceeds to the final stage, the \emph{global model} (Section~\ref{subsec: global}). The \emph{global model} is a state-of-the-art graph convolution neural network trained across the Redshift fleet. 

\subsection{Overview}
\label{subsec: workflow}

\begin{figure}[t]
    \centering
    \includegraphics[width=8cm]{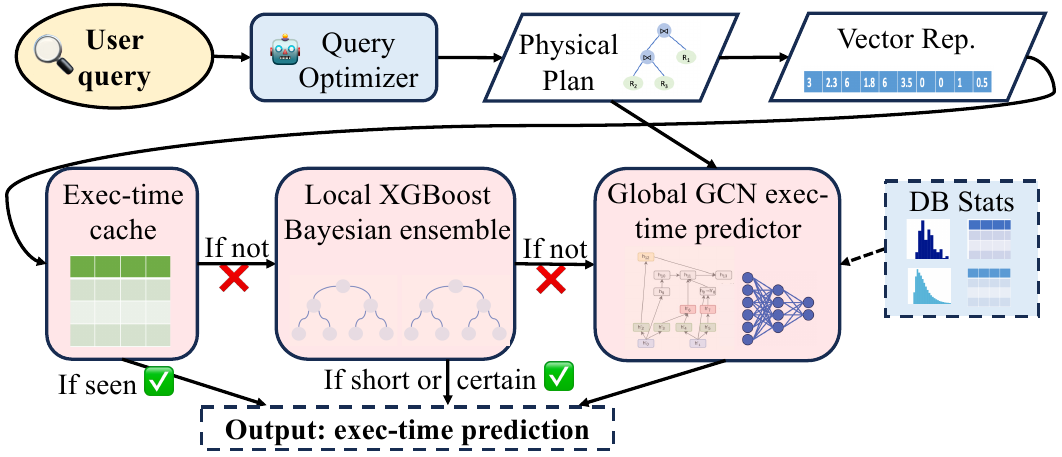}
    \caption{Workflow overview of the \ours predictor: Queries are first planned by the query optimizer and then featurized into a vector. If the exec-time cache has a match for the feature vector, the cached exec-time will be returned. Otherwise, the feature vector will go to the local model. If the local model predicts that a query is short-running or if it is highly confident, its prediction will be returned. Otherwise, the global model will use the original physical plan and stats from the user's specific database to make a prediction. After execution, the pair of vector representation and exec-time is added to the cache and local training data (not shown).}
    \label{fig: stagemodel}
\end{figure}

We present an overview of \ours predictor's workflow in Figure~\ref{fig: stagemodel}. 
As discussed in Section~\ref{sec: background}, a user's query $\mathcal{Q}$ will go through the parser and query optimizer to derive its physical execution plan before arriving at the exec-time predictor. 
\ours predictor first flattens the physical plan of $\mathcal{Q}$ into a 33-dimensional vector. 
The hashed vector is checked against the \textit{exec-time cache}, which records the observed queries and their actual exec-time in the past. 
If $\mathcal{Q}$ is in the cache, we return the prediction based on its observed exec-time (within the Redshift, we find that $60\%$ of the queries will find a match in the cache).
When $\mathcal{Q}$ misses the cache, we send $\mathcal{Q}$'s vector representation to the \textit{local model} that is trained on executed queries on this instance, which will output a prediction and uncertainty associated with it. 
If the \textit{local model} thinks that $\mathcal{Q}$ is short-running, or the local model is highly confident about its prediction, \ours predictor directly returns the local model's prediction.

Finally, when the \textit{local model} is uncertain about $\mathcal{Q}$, the state-of-the-art graph convolution neural network (\textit{global model}) is used.
The \textit{global model} takes the physical plan of $\mathcal{Q}$ as input and understands the complicated interactions between operator nodes in this plan tree. 
A single \textit{global model} is trained on all executed queries from a diverse set of instances and is shared across the entire Amazon Redshift fleet. 
As a result, the \textit{global model} is more robust and accurate on the queries that the \textit{local model} is uncertain about.
The downside to the global model is that the global model has a high inference time (e.g., up to 100ms). 
However, since it is only used when \textit{local model} is uncertain and thinks the query is long-running, it is rarely used, so the additional time is amortized out.

\paragraph{Advantages}
The \ours predictor designs a systematic approach to use these models that best leverages the optimality (\textit{exec-time cache}), instance-optimization (\textit{local model}), and transferrable knowledge about Redshift (\textit{global model}). 
By combining the merits of all these models with different characteristics, the \ours predictor is able to reliably achieve high prediction accuracy at a (amortized) negligible inference latency. 
Conceptually, the \ours predictor effectively addresses the downside of the prior AutoWLM predictor inside Redshift. 
In particular, the \ours predictor is able to significantly improve prediction accuracy without adding too much inference latency. 
In addition, the \ours predictor has reliable performance, especially when used on a new instance with insufficient training queries or instances with changing data and query workload.
For those instances, the uncertainty of \textit{local model}'s prediction will be high, and we can rely on the more robust \textit{global model}.
At last, the \ours predictor can provide probabilistic
distribution or confidence interval for the predicted exec-time to enable robust decisions making of many downstream tasks.

\subsection{Exec-time Cache}
\label{subsec: cache}

The \textit{exec-time cache} is able to output near-optimal prediction with near-zero inference latency for repeating queries that were recently observed (this is $60\%$ of queries on average across the Redshift fleet). In the following, we first explain the cache's keys and values, the procedure to make predictions, and the eviction policy of entries when the cache is full. Then, we discuss several optimizations.

\paragraph{Cache keys and values} 
Similar to AutoWLM predictor~\cite{SaxenaRCLCCMKPN23}, the first step of the \ours predictor is to flatten a physical plan tree as a vector. We traverse the plan tree, collect operator nodes of the same type, and sum up their estimated cost and cardinality. We also add features such as query type (e.g., \texttt{SELECT}, \texttt{DELETE}) and end up with an $n$-dimensional vector representation of the physical plan tree.  
The \textit{exec-time cache} uses this vector for each query as key and maps the actual exec-time of this query after execution as values. For the same query that executes multiple times, the cache stores all their observed exec-times as values.
It is worth noticing that in some very rare cases, two different plan trees may result in the same vector representation, which means that the \textit{exec-time cache} cannot distinguish them. However, in those cases, their query plans should be very close to each other, and thus, we assume their exec-times should be similar as well.

\paragraph{Exec-time prediction}
Whenever a new query arrives that matches a cached key, the \textit{exec-time cache} is able to predict its exec-time based on its observed exec-times $t_1, \ldots, t_k$.
Since variance exists in these observed exec-times due to different system loads when the same query is being executed, one might think to use the mean $\mu$ of these exec-times as a prediction to increase prediction robustness.
However, since the underlying table stats of Redshift may not be up-to-date, the same query executed at different time may access slightly different data, leading to different exec-times. 
In this case, $\mu$ will contain outdated exec-times, and the most recently observed exec-time $t_k$ captures the freshness of data.
Therefore, we design a simple heuristic to predict the exec-time: $\mu \times \alpha + t_k \times (1-\alpha)$.
This heuristic can capture both the robustness and the freshness of data. The value of $\alpha$ is a hyperparameter that balances the average exec-time against the most recently observed exec-time. 
Empirically, $\alpha = 0.8$ works well for the Redshift fleet.
In the future, we plan to design more principled approaches for prediction based on observed exec-time, such as time series prediction.

\paragraph{Eviction policy}
In order to maintain efficient memory usage and fast look-up speed, we need to ensure the number of cached queries does not grow unbounded.
Therefore, whenever the number of cached queries surpasses a certain threshold, \textit{exec-time cache} will evict the least updated queries from the cache. 
In practice, this can be implemented by maintaining a sorted list of dates for each query at which the most recently observed exec-time is collected and removing the query with the oldest date from the \textit{exec-time cache}. 

\paragraph{Optimization 1: hash value replacement} As described above, the hash table stores query feature vectors as keys, so whenever an incoming query arrives, its entire query feature vector needs to be compared element-by-element with the cached vectors. 
We can optimize this vector-vector comparison by storing the hash value of the feature vector as the key. 
This, in theory, may lead to collisions, as any query feature vectors with matching hash values will be treated as identical, but we observed \emph{zero} hash collision for all queries in the top 200 instances in the Amazon Redshift fleet.
This optimization removes the costly vector-vector comparison and significantly reduces the size of the hash table keys.

\paragraph{Optimization 2: running mean and variance} As described above, the values in the hash table are lists of past query latencies. This gives us some design flexibility, as we can compute any summary statistic we want from the history (e.g., mean, median, quantiles). However, if we know we are only interested in the mean, variance, and the most recently observed exec-time, we can replace each query history with a running mean and variance. 
Using Welford's algorithm~\cite{welford}, this only requires storing 4 values per hash table entry, reducing both the in-memory size of the cache and the cache's lookup time (due to reading fewer in-memory values).

\subsection{Instance-optimized local model}
\label{subsec: local}

\comment{
\begin{itemize}
    \item Refer back to background, explain why it makes sense in other scenario (each type of uncertainty and direct corresponding in our settings) 
    \item Explain the how to leverage cache to do data deduplication technique 
    \item discuss its implementation and practicality (i.e. minimal changes to the current model, model size and inference time is reasonably small)
    \item \ziniu{Should we interleave the mini-benchmark experiment here directly? I.E. accuracy is the same as the previous local model, the uncertainty measure is reliable.}
\end{itemize}
}

In this section, we first describe the implementation of our \emph{local model}, and then conceptually justify the design choices for \emph{local model} in Redshift and compare it with other potential alternative designs. 
At last, we explain how to leverage the \emph{exec-time cache} to build an effective and efficient training pipeline for \emph{local model}.

\paragraph{Bayesian ensemble of XGBoost models} As mentioned in Section~\ref{subsec: related}, we adapt and implement the Bayesian ensemble of XGBoost models~\cite{MalininPU21} as our instance-optimized \emph{local model} with a reliable uncertainty measurement. 
Recall that this ensemble independently learns $K$ XGBoost models, each of which takes the 33-dimension vector representation of query as input and estimates a mean $\mu_k$ and variance $\sigma^2_k$ of a query's exec-time.
The final prediction of exec-time $\hat{y}$ is given by the average of each model's prediction in Equation~\ref{equ: pred_mean}. 
The total uncertainty $\mathcal{V}[\hat{y}]$ (i.e., variance of prediction) of this prediction $\hat{y}$ is a summation of estimated model uncertainty and estimated data uncertainty as shown in Equation~\ref{equ: uncertainty}. 

\begin{equation}
\label{equ: pred_mean}
    \hat{y} = \frac{1}{K} \sum_{k=1}^K \mu_k
\end{equation}

\begin{equation}
\label{equ: uncertainty}
    \underbrace{\mathcal{V}[\hat{y}]}_{\text{Prediction uncertainty}} = \ \  \underbrace{\frac{1}{K} \sum_{k=1}^K (\hat{y} - \mu_k) ^ 2}_{\text{Model uncertainty}} \ \ \ +  \ \ \underbrace{\frac{1}{K} \sum_{k=1}^K \sigma^2_k}_{\text{Data uncertainty}}
\end{equation}

\paragraph{Justification of the local model's design choices}
The model uncertainty is estimated as the variance of each XGBoost model's prediction $\mu_k$. 
Since each model is independently trained, when \emph{local model} does not have enough training data or if the incoming query is different from the training queries, the models will have diverse interpretations of this query.
Thus, the variance of their prediction will be high in this scenario, and the \textit{global model} could come to the rescue.

The estimated data uncertainty can capture the inherent noisiness in the labels and training features themselves. 
In Redshift, the same query executed at a different time can have different exec-times (i.e. noisiness in labels) due to different system loads and concurrency state.
Meanwhile, the input to \emph{local model} also contains high noise. Specifically, the 33-dimensional vector feature does not fully capture all information of the physical query plan tree (e.g., tree structure, missing node types, and Redshift's cardinality estimation error). 
When a query plan is complicated with many joins, the vector feature tends to be less representative and the \emph{local model} will have a high data uncertainty. In this case, the \textit{global model} will take the entire physical execution plan as input and will have a better performance. 

Therefore, using the Bayesian ensemble of XGBoost models as the \emph{local model} captures two sources of uncertainty that could result in high prediction errors in Redshift. There exists a line of works in the machine learning domain for quantifying the prediction uncertainty, which are less optimal to apply inside Redshift. 
Specifically, uncertainty measurement using deep learning models~\cite{GalG16, Lakshminarayanan17, WilsonI20} are not practical in Redshift due to their large inference latency. The popular lightweight alternatives for uncertainty measurement normally only focus on one source of uncertainty. 
For example, uncertainty in random forest regression~\cite{MentchH16, coulston2016approximating}, quantile regression forests~\cite{Meinshausen06}, and one-class support vector machine for outlier detection~\cite{li2003improving, amer2013enhancing} mainly focus on quantifying the model uncertainty but not the data uncertainty. Whereas, another line of works on probabilistic prediction~\cite{gneiting2014probabilistic, nowotarski2018recent, DuanADTBNS20, roberts1965probabilistic} and probabilistic programming~\cite{gordon2014probabilistic, bingham2019pyro} mainly focus on understanding the uncertainty in data itself rather than quantifying the model uncertainty.

It is worth noting that using the Bayesian ensemble of XGBoost models as \emph{local model} in Redshift also involves minimal engineering effort since the prior AutoWLM predictor inside Redshift already uses XGBoost. In order to build the new local model, we only need to change the loss function of the XGBoost model and independently train multiple such models.

\paragraph{Local model training optimization} 
The training process of the local model needs to maintain a diverse set of training queries to train effective \textit{local models}. At the same time, it also needs to ensure a low memory and computation overhead of training
because the training process is conducted locally on the customers' database clusters. Therefore, we tailor the training process based on the unique characteristics of Redshift queries.

We collect the observed features and latency of executed queries into a training query pool. Naively storing every query execution result in the training pool has the following three issues: (1) the size of the training pool would grow unbounded, (2) the training pool would become ``polluted'' with repetitive queries that the exec-time cache will take care of anyway, and (3) the training pool will have more short queries than long queries, skewing prediction accuracy for longer (and often more important) queries.

\paragraph{Bounding the size} To resolve the first issue of the training pool, we cap the total number of queries in the training query pool. Whenever the number of queries exceeds a certain threshold, the training pool will evict the oldest observed queries.

\paragraph{Dealing with repeats} Recall that a large amount of queries in Redshift repeat themselves, which will significantly reduce the diversity of queries in the training pool. Besides, these repeating queries will be captured by the \textit{exec-time cache}, so overfitting these queries may degrade \textit{local model}'s generalizability to other queries. 
Therefore, we deduplicate the repeating queries in the training pool. We leverage the \textit{exec-time cache} to implement this data deduplication efficiently. 
Specifically, for each executed query, we hash its observed features and check against the \textit{exec-time cache}. If the query misses the cache, we add it to the training pool.

\paragraph{Duration diversity} Next we address the third issue, which is that the distribution of query latencies is skewed. Most of the Redshift queries execute in less than a couple of seconds, so our training pool can be filled by short-running queries. In this case, the \textit{local model} will have catastrophic performance for longer-running queries. Therefore, we partition the training pool into several query exec-time buckets (e.g., $0-10s$, $10-60s$, and $60s+$) and assign a cap for each bucket to ensure the query diversity in the training pool.

\subsection{Transferrable global model}
\label{subsec: global}

Inspired by recent work in zero-shot cost model~\cite{HilprechtB22}, we design the instance-independent featurization of Redshift query plans, allowing us to map query plans from various customers' instances to a unified space.
As a result, we can collect a diverse set of training queries from a large amount of Redshift instances to jointly train one \textit{global model} that is able to make robust predictions for all instances, including the unseen instances. 
The \textit{global model} uses a graph convolutional network (GCN)~\cite{gcn} architecture to understand the query plan of Redshift and map it to its exec-time.

\begin{figure}[t]
	\centering
	\includegraphics[width=8cm]{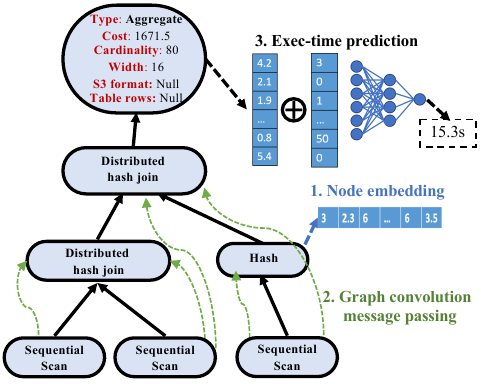}
	\caption{Global model featurization and architecture.}
	\label{fig:global_model}
\end{figure}

\paragraph{Query plan featurization}
We run the fleet sweep to gather the logs (i.e., \texttt{STL\_EXPLAIN} table) on the physical execution plans of executed queries from the customers' Redshift instances.
Then, we parse the information from the logs into a tree data structure representing each query plan as shown in Figure~\ref{fig:global_model}.
Each node in the tree represents a physical operator (e.g., ``sequential scan'', ``hash'', ``materialize'',  ``distributed hash join'', ``aggregate'', ``order by''), and we featurize it as its operator type, estimated cost, estimated cardinality, tuple width, S3 table format (e.g., ``Parquet'', ``OpenCSV'', ``Text'' or ``Local'' if the table is stored locally), and number of rows in the table. An example of node features is shown in red in Figure~\ref{fig:global_model}. It is worth noticing that 90 unique operator types exist in Redshift, so we represent the node operator type as a $90$-bit one-hot vector. Furthermore, we set the S3 table format and table rows features to ``Null'' if the operator is not directly operating on a base table (e.g., not a scan operation).

\paragraph{Model architecture}
Our global exec-time predictor contains three components: node embedding, graph convolution message passing, and final exec-time prediction. 
First, the features of all nodes are embedded with a multi-layer perceptron (MLP) to a feature vector. One example is shown in blue color in Figure~\ref{fig:global_model}.
Then, we use a GCN model to perform message passing between nodes to aggregate information and understand operator interactions in Redshift. 
Some message passing directions are shown in green in Figure~\ref{fig:global_model}. 
Specifically, GCN consists of several layers of message passing.
In the first layer of GCN, each node combines its own embedded node features with those of its children and transforms them into a new node feature. 
This feature combination process is controlled by learnable weights. 
The following GCN layers will repeat the same process on the transformed features of each node from the previous layer.
After several GCN layers, information on all nodes will be aggregated to the root node, and a vector representation of the entire query plan will be outputted. The GCN message passing is capable of understanding the complex operator dependencies and interactions, as shown by the zero-shot cost model~\cite{HilprechtB22}.
At last, the final vector representation output by GCN will be concatenated with a system feature vector, which includes a summarization of the query plan, Redshift instance type, number of Redshift nodes, memory size, and number of concurrent queries. 
The system feature vector contains factors that may affect query exec-time other than the query execution plan itself.
The concatenated feature will be sent to an MLP to estimate the exec-time of this query.

Our global GCN model is trained on a diverse set of hundreds of Redshift instances, each with more than 10,000 queries.

\section{Experimental Evaluation}
\label{sec: exp}
In this section, we evaluate the performance of the \ours predictor and compare it against the prior AutoWLM predictor in Redshift on real-world data. We first explain the experimental setting in Section~\ref{subsec: exp-setting} and then evaluate the following questions:

\begin{itemize}
    \item \textbf{End-to-end (Section~\ref{subsec: exp-2e2}):} How much practical gain can the \ours predictor achieve in terms of improving end-to-end query execution latency in Redshift? 
    \item \textbf{Accuracy (Section~\ref{subsec: exp-acc}):} How accurate is the \ours predictor? What is its model size and inference latency?
    \item \textbf{Ablation (Section~\ref{subsec: exp-ablation}):} How accurate and robust is each hierarchy of \ours predictor individually? 
\end{itemize}

\subsection{Experimental Settings}
\label{subsec: exp-setting}

\paragraph{Real-world workloads}
Since we are primarily concerned with the performance of the \ours predictor on the Redshift fleet, we evaluate the \ours predictor on query logs of real Redshift customers.
We select the top 100 most-billed instances in the month of July 2023 from Redshift in each of the three regions: `us-east-1', `us-west-2', and `eu-west-1'.
We select all user-executed queries on these instances from `July 28th, 2023' to `August 11th, 2023', resulting in a total of roughly 30 million queries.
Unless specifically stated otherwise, all our experiments are conducted on these queries.

\paragraph{Local environment} 
We train and evaluate the performance of our \ours predictor and the baseline (i.e., AutoWLM predictor in Redshift) on one AWS `m5.4xlarge' machine with 64 GB memory, 16 vCPUs, and Intel Xeon Platinum 8175 processor. It is worth noting that the offline training of our \textit{global model} is conducted on two AWS `p3.8xlarge' machines, each with 244 GB memory, 32 vCPUs, and 4 Nvidia tesla v100 GPU. We only leverage GPU to accelerate the training of the \textit{global model}, not the cache or local model.

\paragraph{Model training}
We simulate the exact training and testing procedure as in real Redshift deployment to evaluate the performance of \ours predictor and the baseline. 
Specifically, on each cluster, we replay all the queries sequentially based on their logged execution start time. 
In addition, we randomly sample 100 training instances and use three weeks of user-executed queries on all instances to train our \textit{global model}. These training instances do not overlap with the evaluation instances.
Unless specified otherwise, all predictions used for evaluation are derived from the aforementioned procedure.

\paragraph{Hyper-parameters}
Our \ours predictor contains a set of hyper-parameters that are relatively easy to tune. We describe each of the hyper-parameters and its value as follows. Each value was determined via tuning on data prior to our test workload (i.e., data used to select hyperparameters is completely separate from the evaluation dataset).
For the \textit{exec-time cache}, we set the cache size to 2,000 (i.e., it can only keep 2,000 unique queries before eviction).
For the \textit{local model}, we used the CatBoost Python package~\cite{ProkhorenkovaGV18} to train $10$ XGBoost models independently. 
For each XGBoost model, we set the number of estimators as 200, the max depth as 6, and the number of parallel trees as 1. 
When training the XGBoost models, we randomly choose $20\%$ of the training data as a validation set for early stopping to prevent overfitting.
We note that the AutoWLM predictor baseline in Redshift uses exactly the same hyperparameters for the XGBoost model. 
The differences between our \textit{local model} and the baseline are (1) we train 10 models instead of one; (2) we use a log-likelihood loss function instead of the mean absolute error as used by the baseline. 
Both changes are necessary to provide an uncertainty measure of prediction.
For \textit{global model}, we use a directed GCN with the hidden dimension size 512, 8 layers of graph convolution, and $0.2$ weight dropout ratio.

\subsection{End-to-end Evaluation in Redshift}
\label{subsec: exp-2e2}

The most straightforward and important approach to evaluate the effectiveness of \ours predictor is to test how much it can improve the end-to-end query execution latency inside Redshift. It is worth noticing that query \emph{latency} is different from query \emph{exec-time} --- latency includes the scheduling time and wait time of a query, whereas exec-time excludes those factors. 

\paragraph{End-to-end simulation}
To evaluate the impact of the \ours predictor on end-to-end performance, we simulate the Redshift workload manager~\cite{SaxenaRCLCCMKPN23} using the Redshift team's internal tools. The simulator replays an existing user workload using the \ours predictor. The simulator then computes the expected query latency of each query in the workload. More accurate exec-time predictions will cause the workload manager to make better scheduling decisions and thus improve query latency. It is worth noticing that in this simulation, exec-time prediction accuracy only affects query wait time but not actual query exec-time: that is, the query execution time is assumed to be identical to when the query was actually executed by the customer. This could lead to some simulation inaccuracies since the workload manager's decisions can impact query execution time (i.e., resource allocation). However, we have verified through various other experiments conducted internally by Redshift teams on the workload manager that improvements in the simulation results accurately reflect actual query execution latency in production. 

We chose the workload manager simulation as our end-to-end evaluation for two reasons. 
First, we cannot directly compare the \ours predictor with the AutoWLM predictor in real production environments. This is because users execute their queries once and once only, so the workload manager either uses the \ours predictor or the AutoWLM predictor to get the actual query execution latency. Counterfactually ``replaying'' the workload via a simulator is the only way to measure possible improvements.
Second, other tasks exist in Redshift as sub-routines (e.g., automatic materialized view creation). However, those tasks are not on the critical path of query execution, so they can only indirectly reflect the query execution latency changes. Thus, we cannot explicitly evaluate the benefit of \ours predictor on those tasks.
It is worth noticing that part of the \ours predictor (\textit{exec-time cache} and \textit{local model}) is already deployed in production. Due to some observed regression in the accuracy of \textit{global model} (see Section~\ref{subsec: exp-ablation}), we are exploring a more robust \textit{global model} before deploying it in Redshift. 

\begin{figure}[t]
    \centering
    \includegraphics[width=8cm]{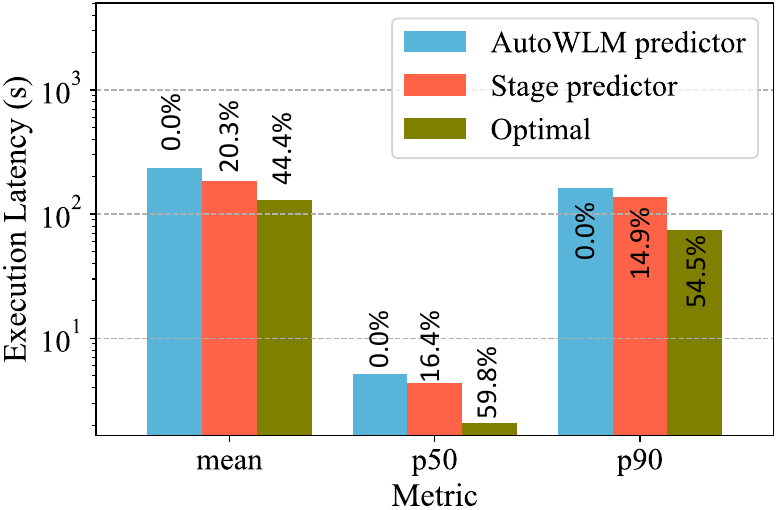}
    \vspace{-1em}
    \caption{End-to-end performance in terms of query latency of different exec-time predictors inside Redshift. We listed the percentage improvement over the AutoWLM predictor.}
    \label{fig:e2e_perf}
\end{figure}

\paragraph{Performance comparison}
We conduct the simulation experiment using three exec-time predictors: \ours predictor, the AutoWLM predictor, and the actual exec-time (\textit{Optimal}). 
The \textit{Optimal} feeds the observed exec-time to the workload manager, representing the optimal performance an exec-time predictor can ever achieve. 

The overall end-to-end query execution latency on all top 100 most-billed instances in three regions (roughly 30 million queries) are shown in Figure~\ref{fig:e2e_perf}.
We observe that the \ours predictor significantly improves over the AutoWLM predictor: the $20.3\%$, $16.4\%$, and $14.9\%$ query execution latency improvement on average, median, and tail, respectively.  
This improvement purely results from a more accurate exec-time predictor.
However, we observe that \textit{Optimal} has a significantly better performance than \ours predictor: $44.4\%$, $59.8\%$, and $54.5\%$ query execution latency improvement on average, median, and tail, respectively.
This suggests that there still exists a large room for improvement, and further improving the accuracy of exec-time predictor in Redshift can be fruitful.

\begin{figure}[t]
    \centering
    \includegraphics[width=7.5cm]{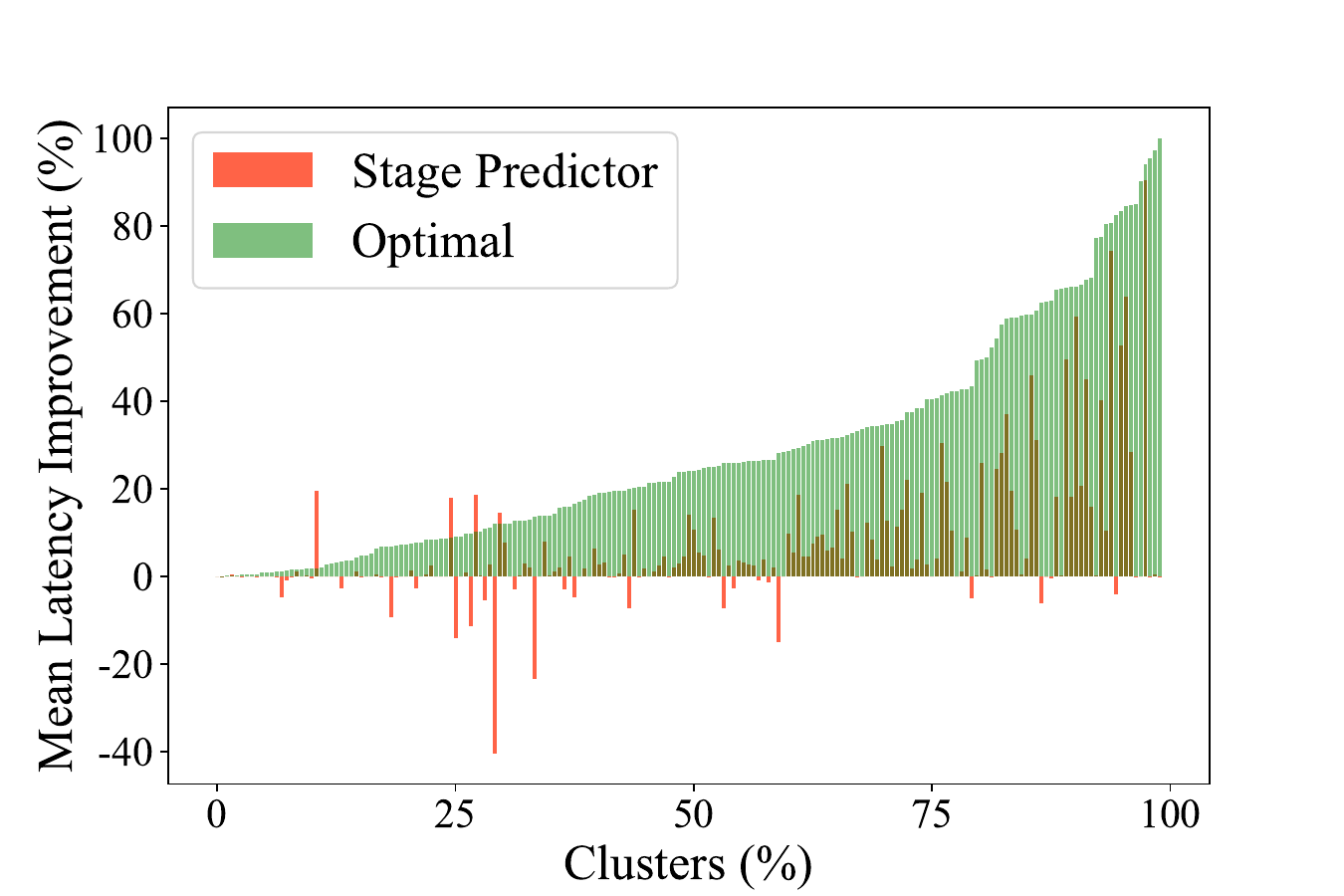}
    \vspace{-1em}
    \caption{End-to-end query latency improvement over AutoWLM predictor on each top instance. We sort the instances based on the improvement the optimal predictor achieves.}
    \label{fig:e2e_instances}
\end{figure}

In addition, we analyze the average query latency improvement of the \ours predictor over the AutoWLM predictor on each instance in Figure~\ref{fig:e2e_instances}. For comparison, we also plot the optimal predictor's improvement and sort the instances based on this value. 
We can see that the \ours predictor is able to improve the average query latency for most of the instances. 
However, there are regressions: on less than $10\%$ of the instances, \ours predictor actually does worse than the AutoWLM predictor. 
There are several possible explanations.
First, the AutoWLM predictor does occasionally make better predictions than the \ours predictor on a small portion of queries, which could have an impact on the workload manager. 
Second, both the \ours predictor and the AutoWLM predictor have erroneous predictions, as detailed in Section~\ref{subsec: exp-acc}, and there is an asymmetry in prediction errors. 
For example, for a query with exec-time of 30s, the \ours predictor may predict it to be 5s, thus sending it to the short-running queue, and the AutoWLM predictor may make a worse prediction of 900s, thus sending it correctly to the long-running queue.
In this scenario, although \ours predictor is more accurate, it makes the wrong decision. 
Third, due to the algorithmic design of workload manager~\cite{SaxenaRCLCCMKPN23}, there will be edge cases when perfect prediction does not lead to the best end-to-end query latency. This also explains why the \ours predictor can sometimes outperform the optimal predictor.

\subsection{Stage Predictor Accuracy}
\label{subsec: exp-acc}

\begin{figure}[t]
	\centering
	\includegraphics[width=7.5cm]{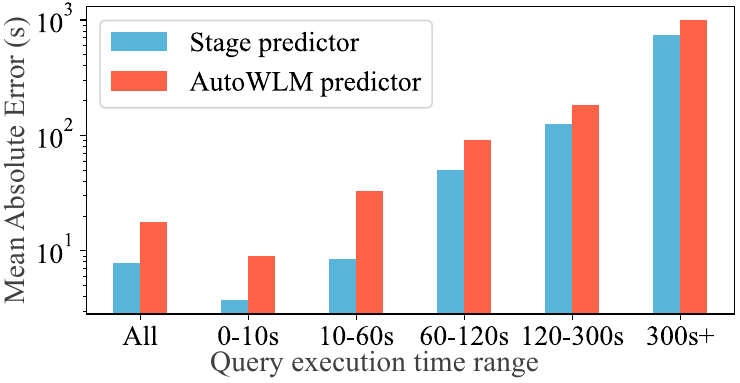}
	\vspace{-1em}
	\caption{Prediction accuracy of stage predictor compared to the AutoWLM predictor in Redshift.}
	\label{fig: overall_ae}
\end{figure}

We show the prediction accuracy of the \ours predictor and the AutoWLM predictor on all top 100 most billed instances from three regions in Redshift, with a total of $27,441,359$ queries.
The accuracy is evaluated using absolute error, that is, |\textit{actual exec-time} $-$ \textit{predicted exec-time}| in seconds.
We show the accuracy comparison in Figure~\ref{fig: overall_ae} and report the details of mean (MAE), median (p50-AE), and tail (p90-AE) absolute error of these queries in Table~\ref{tab: overall-ae}.
\ours predictor is able to achieve a median absolute error of $0.67$, suggesting that for $50\%$ of the query, \ours prediction is within $0.67s$ of actual execution time.
Overall, \ours predictor achieves more than 2x more accurate prediction than the AutoWLM predictor. 

\begin{table}[t]
	\scalebox{0.75}{
		\begin{tabular}{c|c|ccc|ccc}
			\hline
			\multirow{2}{*}{Query Exec-time} & \multirow{2}{*}{\# Queries} &
			\multicolumn{3}{c}{\textbf{Stage predictor}}  & \multicolumn{3}{c}{AutoWLM predictor} \\ \cline{3-8}
			& & MAE & P50-AE & P90-AE & MAE & P50-AE & P90-AE \\ \hline
			Overall & 27,441,359 & 7.76 & 0.67 & 9.39 & 17.87 & 2.03 & 23.68 \\ \hline 
			0s -- 10s & 22,015,851 & 3.74 & 0.31 & 7.43 & 9.04 & 1.11 & 14.44 \\ \hline 
			10s -- 60s & 5,085,965 & 8.53 & 2.60 & 13.68 & 32.83 & 10.02 & 51.34 \\ \hline 
			60s -- 120s & 163,913 & 50.11 & 24.15 & 85.00 & 91.63 & 31.04 & 113.8 \\ \hline 
			120s -- 300s & 83,590 & 126.4 & 70.46 & 206.4 & 181.9 & 84.55 & 255.4 \\ \hline 
			300s+ & 92,041 & 744.4 & 235.7 & 1496 & 990.1 & 289.7 & 1922 \\ \hline 
		\end{tabular}
	}
	\caption{Prediction accuracy (absolute error in seconds) of stage predictor and the AutoWLM predictor.}
	\vspace{-2em}
	\label{tab: overall-ae}
\end{table}

\begin{table}[t]
	\scalebox{0.75}{
		\begin{tabular}{c|c|ccc|ccc}
			\hline
			\multirow{2}{*}{Query Exec-time} & \multirow{2}{*}{\# Queries} &
			\multicolumn{3}{c}{\textbf{Stage predictor}}  & \multicolumn{3}{c}{AutoWLM predictor} \\ \cline{3-8}
			& & MQE & P50-QE & P90-QE & MQE & P50-QE & P90-QE \\ \hline
			Overall & 27,441,359 & 54.57 & 1.60 & 19.00 & 171.8 & 4.08 & 135.7 \\ \hline 
			0s -- 10s & 22,015,851 & 43.87 & 1.92 & 26.25 & 97.41 & 6.38 & 173.1 \\ \hline 
			10s -- 60s & 5,085,965 & 71.66 & 1.18 & 2.16 & 441.4 & 1.77 & 6.39 \\ \hline 
			60s -- 120s & 163,913 & 251.9 & 1.38 & 4.57 & 633.2 & 1.51 & 4.97 \\ \hline 
			120s -- 300s & 83,590 & 307.4 & 1.59 & 5.83 & 548.1 & 1.71 & 6.61 \\ \hline 
			300s+ & 92,041 & 1084 & 1.48 & 6.12 & 1922 & 1.58 & 10.00 \\ \hline 
		\end{tabular}
	}
	\caption{Prediction accuracy (in Q-error) of stage predictor compared to the AutoWLM predictor.}
	\label{tab: overall-qe}
\end{table}

To dive deep into the results, we provide a detailed accuracy comparison on queries with different exec-time ranges in Table~\ref{tab: overall-ae}.
It is very important to analyze the prediction performance of queries with different exec-time ranges because the workload manager of Redshift schedules queries into execution queues and assigns priority according to the predicted exec-time.
Specifically, we see that \ours predictor is able to achieve more than $3x$ better prediction on queries with less than 60s exec-time.
We additionally provide the same table on another widely-used metric for relative error: Q-Error~\cite{bound_card}, that is, $max\{predicted/true, true/predicted\}$ in Table~\ref{tab: overall-qe}. The minimal value Q-Error can take is 1, and closer to 1 implies a more accurate prediction.
We observe a roughly similar pattern that the \ours predictor significantly outperforms the AutoWLM predictor on 60s exec-time. However, it achieves mild improvement on the queries with more than 60s exec-time, possibly due to the following reasons.
First, the distribution of query exec-time is heavily skewed and only $1\%$ of the queries execute longer than 60s.
Therefore, both \ours predictor and the AutoWLM predictor don't have enough training data on the long-running queries and yield worse performance. 
Second, the long-running queries are inherently more difficult to predict because of a larger noisiness in the label. 
We observe in Redshift that the exec-time of the same query repeatedly executed several times can range from just tens of seconds to several hundred seconds. 
This scenario is largely due to different system loads, cache states, and the number of concurrency queries in Redshift during the query execution.
Unfortunately, our exec-time predictor cannot yet take that into account.
In future work, we plan to design better exec-time predictors that can take system statistics into account to explain the variance in labels.

\begin{figure}[t]
    \centering
    \includegraphics[width=8cm]{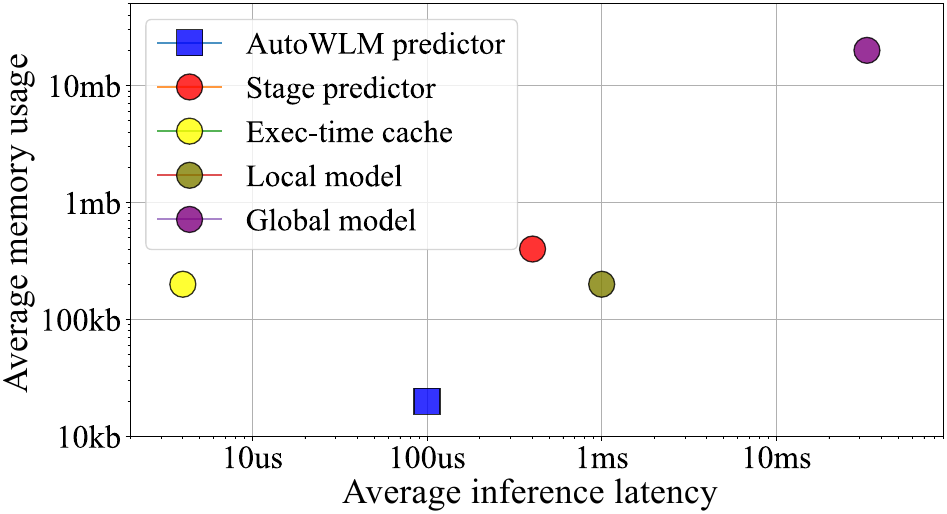}
    \vspace{-1em}
    \caption{Average inference latency and average memory usage overhead of different exec-time predictors.}
    \vspace{-1em}
    \label{fig:inference}
\end{figure}

In addition to accuracy, we provide the average inference latency and memory usage overhead of \ours and AutoWLM predictor along with each component of \ours predictor in Figure~\ref{fig:inference}. 
It is worth noticing that since different instances of Redshift are likely to have different hardware types, the numbers in Figure~\ref{fig:inference} are rough estimations rather than actually measured inference latency and memory usage.
Although \ours predictor (red dot) has a larger inference latency and memory overhead than the AutoWLM predictor (blue square), it is still within a practical range: sub-millisecond inference latency, a few hundred kb memory usages.
Specifically, the \textit{exec-time cache} (yellow dot) is able to make an inference in just a couple of microseconds. 
The \textit{local model} (green dot) trains $10$ XGBoost models as opposed to one in AutoWLM predictor, so it is generally $10x$ larger and slower to make inferences than AutoWLM predictor.
The \textit{global model} (purple dot) is a deep-learning-based model, which is roughly two of a magnitude larger than other predictors. 
However, since the deep learning model is rarely used ($3\%$ of the time), its inference latency is amortized out. 
Furthermore, we do not include the memory overhead of \textit{global model} into \ours predictor because it will eventually be deployed as a serverless Lambda function that every Redshift instance can invoke to avoid local memory and CPU overhead.

\subsection{Ablation study}
\label{subsec: exp-ablation}

In this section, we first provide detailed accuracy analysis on each component of \ours predictor: \textit{exec-time cache}, \textit{local model}, and \textit{global model}. 
Then, we study how reliable is the uncertainty measure of our \textit{local model}.

\paragraph{Accuracy of exec-time cache}
\begin{table}[t]
	\scalebox{0.75}{
		\begin{tabular}{c|c|ccc|ccc}
			\hline
			\multirow{2}{*}{Query Exec-time} & \multirow{2}{*}{\# Queries} &
			\multicolumn{3}{c}{\textbf{Exec-time cache}}  & \multicolumn{3}{c}{AutoWLM predictor} \\ \cline{3-8}
			 & & MAE & P50-AE & P90-AE & MAE & P50-AE & P90-AE \\ \hline
			 Overall & 16,963,658 & 4.83 & 0.56 & 4.66 & 15.04 & 3.00 & 20.35 \\ \hline 
			 0s -- 10s & 12,616,915 & 1.82 & 0.16 & 2.74 & 7.40 & 1.20 & 15.96 \\ \hline 
			 10s -- 60s & 4,212,128 & 4.55 & 2.27 & 6.61 & 23.15 & 7.57 & 24.26 \\ \hline 
			 60s -- 120s & 74,604 & 30.87 & 9.80 & 67.93 & 43.88 & 23.29 & 77.2 \\ \hline 
			 120s -- 300s & 27,185 & 115.8 & 72.78 & 197.8 & 117.4 & 88.13 & 205.6 \\ \hline 
			 300s+ & 32,826 & 764.3 & 193.4 & 1524 & 1046 & 284.4 & 2045 \\ \hline 
		\end{tabular}
	}
	\caption{Prediction accuracy (absolute error in seconds) of exec-time cache and AutoWLM predictor in Redshift.}
    \vspace{-1em}
	\label{tab: acc-cache}
\end{table}

We find that $16,963,658$ out of the $27,441,359$ queries in these instances ($61.8\%$) repeat themselves and can be directly predicted by \textit{exec-time cache}.
As shown in Table~\ref{tab: acc-cache},
\textit{exec-time cache} overall achieves a significantly better prediction accuracy than the AutoWLM predictor. 
The advantages are apparent since the AutoWLM predictor (XGBoost model) is trained on the executed queries' exec-time as ground truth, which is captured in the cache. Therefore, in theory, the locally trained model can never outperform the \textit{exec-time cache}.
However, it is worth noticing that \textit{exec-time cache} does make significant errors (in terms of absolute error) because these repeating queries are executed at different system loads, buffer pool states, and concurrency conditions, making it extremely hard to predict the exec-time at a different state accurately. 
We find roughly the same pattern for prediction accuracy comparison in terms of Q-error, so we omit all results on Q-error due to space limitations.

\begin{table}[t]
	\scalebox{0.75}{
		\begin{tabular}{c|c|ccc|ccc}
			\hline
			\multirow{2}{*}{Query Exec-time} & \multirow{2}{*}{\# Queries} &
			\multicolumn{3}{c}{\textbf{Local model}}  & \multicolumn{3}{c}{AutoWLM predictor} \\ \cline{3-8}
			 & & MAE & P50-AE & P90-AE & MAE & P50-AE & P90-AE \\ \hline
			 Overall & 10,477,701 & 21.48 & 4.16 & 34.88 & 19.06 & 4.32 & 29.27 \\ \hline 
			 0s -- 10s & 9,398,936 & 13.76 & 3.27 & 30.88 & 10.94 & 3.60 & 23.41 \\ \hline 
			 10s -- 60s & 873,837 & 35.63 & 16.47 & 87.06 & 32.63 & 12.90 & 54.36 \\ \hline 
			 60s -- 120s & 89,309 & 77.69 & 37.61 & 137.6 & 69.65 & 36.04 & 93.40 \\ \hline 
			 120s -- 300s & 56,405 & 140.1 & 72.03 & 230.5 & 120.7 & 76.75 & 195.1 \\ \hline 
			 300s+ & 59,215 & 840.3 & 267.6 & 1652 & 852.3 & 276.0 & 1729 \\ \hline 
		\end{tabular}
	}
	\caption{Prediction accuracy (absolute error in seconds) of the local model and AutoWLM predictor in Redshift.}
	\label{tab: acc-local}
\end{table}

\paragraph{Accuracy of local model}
We evaluate and compare the performance of the \textit{local model} to the AutoWLM predictor in Redshift on the $10,477,701$ out of the $27,441,359$ queries that miss the \textit{exec-time cache} ($38.2\%$).
Recall that there are only two differences between \textit{local model} and AutoWLM predictor: 1) \textit{local model} independently trains $10$ XGBoost model whereas AutoWLM predictor only uses one; 2) the XGBoost model in \textit{local model} is trained with log-likelihood loss whereas AutoWLM predictor is trained with the absolute error.
We observe in Table~\ref{tab: acc-local} that \textit{local model} is slightly worse than the AutoWLM predictor because the AutoWLM predictor is directly trained to optimize the absolute error, which is the evaluation metric.
As a future work, we plan to lower the gap in performance difference between the two by adding an XGBoost model trained with absolute error into the Bayesian ensemble of XGBoost models in \textit{local model}.

\paragraph{Accuracy of global model}
The \textit{global model} is trained on a diverse set of instances and evaluated on the top-billed instances with unseen queries.
We first compare the performance of \textit{global model} against the \textit{local model} on all queries that miss the cache in Table~\ref{tab: acc-global}.
We observe that \textbf{the \textit{local model} has a better performance than \textit{global model},} especially on long-running queries.
This was surprising to us because it runs against the common wisdom that ``more data makes a better model'': in this case, the \textit{global model} is trained on \emph{significantly} more data than the \textit{local model}. However, the \textit{local model}'s data is much closer in distribution to the test data, and the \textit{local model} is able to win out. While there are some specific databases where the \textit{global model} outperforms the \textit{local model}, the overall trend favors the \textit{local model}, which might be evidence that, in the context of query performance prediction, \emph{better data beats bigger data}. In our opinion, this casts serious doubt on the premise that cloud database operators can train effective ``cross-customer'' models that are better than instance-optimized models.

\begin{table}[t]
	\scalebox{0.75}{
		\begin{tabular}{c|c|ccc|ccc}
			\hline
			\multirow{2}{*}{Query Exec-time} & \multirow{2}{*}{\# Queries} &
			\multicolumn{3}{c}{\textbf{Global model}}  & \multicolumn{3}{c}{\textbf{Local model}} \\ \cline{3-8}
			 & & MAE & P50-AE & P90-AE & MAE & P50-AE & P90-AE \\ \hline
			 Overall & 10,477,701 & 23.82 & 6.42 & 29.53 & 21.48 & 4.16 & 34.88 \\ \hline 
			 0s -- 10s & 9,398,936 & 12.61 & 8.48 & 20.96 & 13.76 & 3.27 & 30.88 \\ \hline 
			 10s -- 60s & 873,837 & 39.42 & 17.92 & 67.58 & 35.63 & 16.47 & 87.06 \\ \hline 
			 60s -- 120s & 89,309 & 166.1 & 111.9 & 204.6 & 77.69 & 37.61 & 137.6 \\ \hline 
			 120s -- 300s & 56,405 & 366.5 & 243.9 & 519.9 & 140.1 & 72.03 & 230.5 \\ \hline 
			 300s+ & 59,215 & 1763 & 701.5 & 3540 & 840.3 & 267.6 & 1652 \\ \hline 
		\end{tabular}
	}
	\caption{Prediction accuracy (absolute error in seconds) of the global model compared to the local model on all queries that miss the exec-time cache.}
  \vspace{-1em}
	\label{tab: acc-global}
\end{table}

\begin{table}[t]
	\scalebox{0.75}{
		\begin{tabular}{c|c|ccc|ccc}
			\hline
			\multirow{2}{*}{Query Exec-time} & \multirow{2}{*}{\# Queries} &
			\multicolumn{3}{c}{\textbf{Global model}}  & \multicolumn{3}{c}{\textbf{Local model}} \\ \cline{3-8}
			 & & MAE & P50-AE & P90-AE & MAE & P50-AE & P90-AE \\ \hline
			 Overall & 361,752 & 134.8 & 10.09 & 164.1 & 164.7 & 25.21 & 196.8 \\ \hline 
			 0s -- 10s & 167,617 & 9.45 & 3.90 & 20.37 & 50.77 & 18.82 & 113.4 \\ \hline 
			 10s -- 60s & 91,714 & 33.31 & 48.46 & 67.58 & 62.12 & 20.14 & 113.0 \\ \hline 
			 60s -- 120s & 37,539 & 64.72 & 35.78 & 92.89 & 82.03 & 34.27 & 151.2 \\ \hline 
			 120s -- 300s & 31,508 & 230.7 & 90.51 & 309.9 & 235.5 & 82.14 & 326.6 \\ \hline 
			 300s+ & 33,343 & 1033 & 423.8 & 2011 & 1046 & 391.9 & 1712 \\ \hline 
		\end{tabular}
	}
	\caption{Prediction accuracy (absolute error in seconds) of  global model compared to local model on uncertain queries.}
 \vspace{-1em}
	\label{tab: acc-global-uncert}
\end{table}

One possible explanation for the relatively poor performance of the \textit{global model} is a lack of model capacity: could a sufficiently large model learn the latent information hidden in each database instance with enough training data? While we are unable to answer this question conclusively, we did find several examples of nearly identical query plans with nearly identical cost and cardinality estimates from different customers with drastically different performances. No amount of data can resolve this issue, as there are two nearly identical training inputs with wildly different desired outputs. Thus, database-specific features may be needed for a global model to learn to differentiate between these pairs.

For the context of this work, we primarily care about the performance of the global model \emph{when the local model is uncertain and thinks the query is long-running}. We evaluate those queries in Table~\ref{tab: acc-global-uncert}. In this scenario, \textit{global model} is able to achieve a better result than the local model. 
This suggests that \textit{global model} is able to provide more robust and reliable prediction whenever the \textit{local model} is uncertain.
However, we barely observe improvement on long-running queries (larger than 120s) over \textit{local model} because those queries are very sparse in each instance and may contain instance-specific characteristics that the \textit{global model} cannot understand.
Overall, we do see a significant performance drop for \textit{local model} itself, shifting from testing on all queries to only uncertain ones. 
This implies that our uncertainty measurement for \textit{local model} is reliable. 
We provide a more exhaustive analysis of the reliability of this uncertainty measurement as follows.

\paragraph{Uncertainty measurement in local model}

\begin{figure}[t]
    \centering
    \includegraphics[width=8cm]{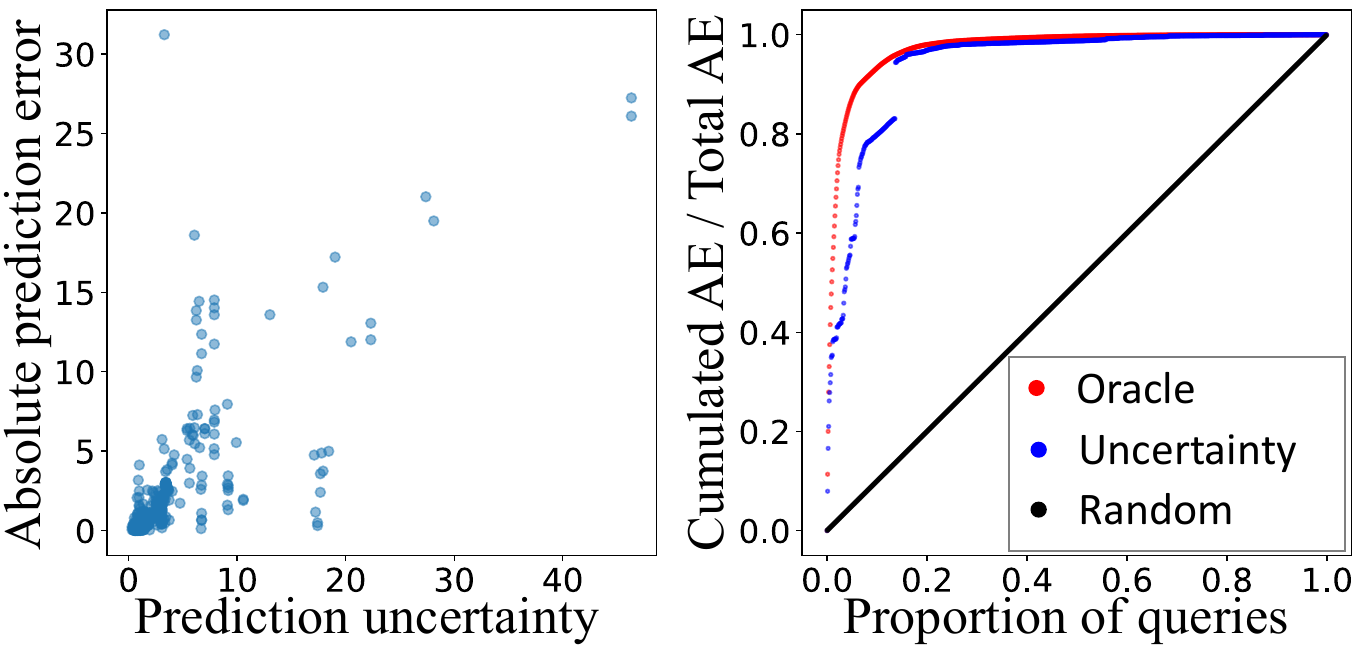}
    	\vspace{-1em}
    \caption{Illustration of PRR calculation on queries from an example instance, whose PPR score is 0.9.}
    \label{fig:prr_example}
\end{figure}

We use the well-\break established scoring rule: prediction-rejection ratio (PRR)~\cite{MalininPU21, MalininMG20} to evaluate how good is the uncertainty measurement of \textit{local model}.
PRR quantifies the rank correlation between the predicted uncertainty and the observed prediction error for each query in one instance.
To better explain how PRR works, we provide an example in Figure~\ref{fig:prr_example}. 
On the left of Figure~\ref{fig:prr_example}, we show 2000 testing queries from one instance and plot their uncertainty estimated by the \emph{local model} on the x-axis against the observed absolute estimation error on the y-axis. 
We can see a statistically significant positive relation between these two values. 
On the right of Figure~\ref{fig:prr_example}, we show the calculation of PRR for this instance.
Specifically, we first sort/rank all queries based on their observed absolute estimation error (``Oracle'') in descending order and plot the proportion of cumulative error (i.e., cumulative error/total error) as shown in the red curve in this figure.
Then, we sort/rank all queries based on their prediction uncertainty in the blue curve and randomly sort all queries as in the black curve. 
Ideally, if our uncertainty has a perfect correlation with the actual error, the blue curve should overlap with the red curve.
Therefore, the ``closer'' blue curve is to the red curve, the more reliable our prediction uncertainty is.
We can compute the area under the curve (AUC) between red and black curves as (AUC\_Oracle) and the AUC between blue and black curves as (AUC\_Stage). The PRR score is quantitatively defined as the ratio AUC\_Stage/AUC\_Oracle, between 0 and 1.

\begin{figure}[t]
    \centering
    \includegraphics[width=7cm]{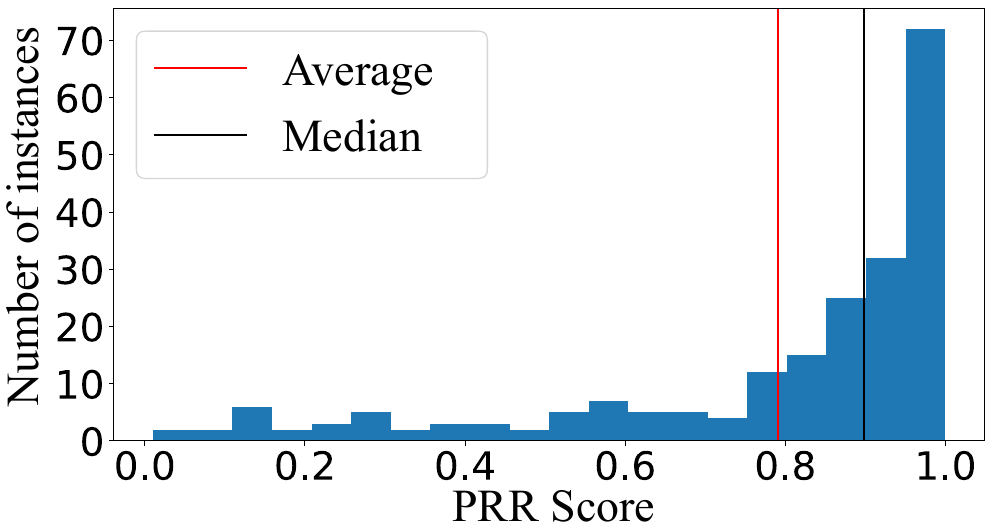}
   	\vspace{-1em}
    \caption{Uncertainty quality of local model (prediction rejection ratio) distribution for all top instances.}
    \label{fig:prr}
\end{figure}

We calculate the overall PRR score for all top instances and plot their distribution in Figure~\ref{fig:prr}.
We can see that $30\%$ of the instance has a PRR score close to 1, which suggests that our uncertainty measurement can perfectly capture estimation error for these instances. 
We have a median PRR score of $0.9$, which is the same as the example in Figure~\ref{fig:prr_example}.
However, in some instances, the PRR score is very low, generally because of insufficient training queries.


\section{Lessons learned and potential future research directions }
\label{sec: lessons}

Throughout the process of building \ours predictor in Redshift, we learned several important lessons that could benefit the research community. In the following, we describe these lessons and list potential research directions that could be valuable to the industry.

\subsection{Applying \ours predictor in other tasks}
Apart from improving the workload manager in Redshift, we believe the \ours predictor enables new solutions for many tasks in smart DBMSes, including query optimization and hypothetical reasoning. 

\paragraph{Query optimization} 
Recently, many ML-based solutions have been proposed for query optimization~\cite{neo, balsa}. 
Although these ML-based techniques can select join-order and physical operators more accurately than the traditional optimizers, their inference latency can get up to several hundred milliseconds. Thus, this overhead will be impractical to optimize short-running queries. 
To practically integrate ML-based optimizers into real systems, they can use \ours predictor as a sub-routine. 
Specifically, a query could first be optimized by the default query optimizer inside a system. Then, the query plan is fed into \ours predictor to estimate its exec-time, and the system will use the expensive ML-based optimizer to re-optimize the plan only if it is long-running. 

\paragraph{Answering ``what-if'' questions} 
Hypothetical reasoning is a crucial element of many database decision-making tasks, including provenance updates, view manipulation, knob tuning, and automatic cluster scaling~\cite{ArabG17, NievaSS20, MeliouS12}. 
Hypothetical reasoning allows DB administrators and users to test database assumptions by asking ``what-if'' questions, such as ``what will the performance of existing queries be if an index on column X is created?'', ``what if the data size increases by 5x?'', ``what if the cluster adds 3 nodes?''.
Answering ``what-if'' questions is very difficult.  For example, learning a query performance predictor on the executed workloads of a database cannot accurately estimate the performance under ``what-if'' scenarios because the model does not observe any training data under such hypothetical scenarios. 
Therefore, existing methods mainly rely on casual inference to answer these questions~\cite{MeliouGMS10, GalhotraGRS22}. 
In theory, the global model of \ours predictor could provide more accurate and more fine-grained answers to these ``what-if'' questions. 
Since the transferrable global model distills the knowledge of a DBMS, it will observe these ``what-if'' scenarios happening on other similar databases. Thus, it can leverage the observation on other databases to accurately predict the query performance under ``what-if'' scenarios of the current database.

\subsection{Hierarchical models}
When designing a practical exec-time predictor for Redshift, we found that although a plethora of ML-based predictors can provide more accurate exec-time prediction, their inference overhead is too large to be deployed on the critical path of Redshift. 
We believe this problem generally exists in the database research community beyond exec-time prediction and Redshift. 

Most ML models naturally present a trade-off between accuracy and model size/inference latency that more accurate models tend to be more expensive. 
Thus, when sophisticated ML-based solutions are adopted to solve existing database problems on the critical path of query execution, they will inevitably incur a non-trivial overhead. 
This overhead may be unaffordable for short-running queries.
We believe the hierarchical model solutions, similar to the \ours predictor, could enable a practical adoption of ML-based solutions to best leverage their accuracy with an affordable inference overhead. 
To the best of our knowledge, there does not exist other works in the database community that use the idea of hierarchical models.
In the following, we provide a detailed example of cardinality estimation, which is on the critical path of query execution.

Cardinality estimation is crucial for query optimization. Due to its challenging nature, sophisticated ML-based solutions~\cite{ZhuWHZPQZC21, HilprechtSKMKB20, YangKLLDCS20} have been proposed to improve the accuracy of their traditional counterpart. Their inference latency varies from a couple of milliseconds to a hundred milliseconds~\cite{HanWWZYTZCQPQZL21}, which will not be affordable for short-running queries. 
A hierarchy of several cardinality estimators with different accuracy/overhead trade-offs could enable practical integration of ML-based solutions in real systems.
Specifically, the queries will first be fed into cheap estimators and more expensive estimators will be invoked only if the previous cheaper estimator is uncertain about its prediction. Therefore, the inference overhead of the expensive estimators can be amortized out.

\subsection{Environment factors in exec-time prediction}
Inside Redshift, we found that the same query in the same cluster can sometimes have very different exec-times ranging from a couple of seconds to several minutes, even hours because of different environment factors at the time of execution. 
These factors include memory and CPU utilizations that directly affect the query exec-time. For example, if $90\%$ of memory has been used by other jobs in the clusters, a query may spill its intermediate results to disk, incurring a large additional cost. However, simply adding the memory and CPU utilizations at the time of execution into the feature of predictor is unlikely to provide better prediction because they can vary throughout the execution of a query.

Furthermore, there exist other environment factors, such as cache effect and buffer pool state that are not trivial to featurize in \ours predictor or any other exec-time predictors.
Specifically, the recently accessed pages will be cached which can greatly speed up the following queries touching the same pages. 

We believe designing exec-time predictors that can accurately take these environment factors into consideration can further improve the prediction accuracy.

\section{Conclusions}
\label{sec: conclusion}

We have presented Stage, a novel hierarchical query performance predictor custom-tailored to Amazon Redshift's specific requirements. The Stage predictor provides fast and robust query performance predictions by taking advantage of the repetitive nature of analytic workloads, remembering the latency of common, frequently-issued queries, and using two different machine learning models for similar and novel queries, respectively. We showed that \ours predictor improves the average query latency by 20\% when compared to Redshift's prior exec-time predictor. 
Based on the lessons learned from building \ours predictor, we pointed out a list of research directions that could be fruitful.

\bibliographystyle{ACM-Reference-Format}
\bibliography{main}

\end{document}